\documentclass{ws-procs9x6}
\usepackage{epsfig}
\begin{document}

\title{Models of Meson-Baryon Reactions 
in the Nucleon Resonance Region }

\author{T.-S. H. Lee$^1$, A. Matsuyama$^2$, T. Sato$^3$ }

\address{$1$ Physics Division, Argonne National Laboratory 
Argonne, IL 60439, USA\\ 
$2$ Department of Physics, Schizuoka University, Schizuoka, Japan \\
$3$ Department of Physics, Osaka University, Osaka, Japan }

\maketitle

\abstracts{
It is shown that most of the models for analyzing meson-baryon reactions
in the nucleon resonance region  can be derived from
a Hamiltonian formulation of the problem. 
An extension of the coupled-channel approach to include $\pi\pi N$ channel
is briefly described and some preliminary results for
the $N^*(1535)$ excitation are presented. }

\section{Introduction}

With very successful experimental efforts in the past few years,
we are now facing a challenge to  interpret very extensive data of 
electromagnetic meson 
production reactions in terms of the structure of 
nucleon resonances ($N^*$).
To achieve this goal, we need to
 perform amplitude analyses of the data in order to
extract $N^*$ parameters. We also need to develop reaction
models to analyze the dynamical content of
 the extracted $N^*$ parameters. At the present time,
we can use the $N^*$ data to
test the predictions from various QCD-based hadron models. In the near future,
we hope to understand $N^*$ parameters from Lattice QCD.

In the $\Delta$ region, both the amplitude analyses and dynamical
reaction models have been well developed.
We find that these two efforts are complementary.
For example, the $\gamma N \rightarrow \Delta$ M1 transition
 strength  extracted from all amplitude analyses is 
$G_M(0) = 3.18 \pm 0.04$ which 
is about 40 $\%$ larger than the constituent quark model prediction.
This difference is understood\cite{sl,dmt}
 by developing  dynamical reaction
models within which one
can show that the discrepancy is due to the pion cloud which is not
included in the commonly considered constituent quark model prediction.
                                                                                
In the second and third resonance regions, the situation is much more
complicated because of many open channels.
It is necessary to develop coupled-channel approaches for learning about
the $N^*$ properties. The main 
objective of this contribution is to
review the development in this direction. 
We will also describe a newly developed
coupled-channel model which is aimed at accounting
 for rigorously the $\pi\pi N$ unitarity condition.

In section 2, we will introduce  
a Hamiltonian formulation within which most of
the current models of electromagnetic
meson production reactions can be derived and
compared. The extension of the coupled-channel approach to account for
explicitly the $\pi\pi N$ channel is then described in section 3.
A summary is given in section 4.

\section{Derivation of Models}

\begin{figure}[t]
\vspace{25pt}
\begin{center}
\mbox{\psfig{file=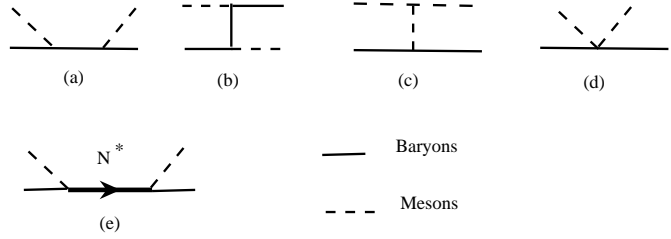, width=100mm}}
\end{center}
\caption[]{ Tree diagrams for meson-baryon reactions. 
$N^*$ is a nucleon resonance state.  }
\label{fig:mechanism_1} 
\end{figure}

The starting point of our derivation is 
to assume that the meson-baryon ($MB)$ reactions can be described by a
Hamiltonian of the following form  
\begin{eqnarray}
H= H_0 + V \,,
\end{eqnarray}
where $H_0$ is the free Hamiltonian and 
\begin{eqnarray}
V = v^{bg} + v^R \,.
\end{eqnarray}
Here $v^{bg}$ is the non-resonant(background) term due to the mechanisms such 
as the tree-diagram mechanisms illustrated in
 Fig.~\ref{fig:mechanism_1}(a)-(d), and $v^R$ describes
 the $N^*$ excitation Fig.~\ref{fig:mechanism_1}(e).
Schematically, the resonant term can be written as
\begin{eqnarray}
v^R(E) = \sum_{N^*_i}\frac{\Gamma^\dagger_i \Gamma_i}{E-M^{0}_i}\,,
\end{eqnarray}
where $\Gamma_i$ defines the decay of the $i$-th $N^*$ state into meson-baryon
states, and $M^0_i$ is a mass parameter related to the resonance position.

The next step is to define a channel space spanned by
the considered  meson-baryon ($MB$)
 channels: $\gamma N$, $\pi N$, $\eta N$, $\pi\Delta$, $\rho N$
$\sigma N$, $\cdot\cdot$. The S-matrix of the 
meson-baryon  reaction is defined by
\begin{eqnarray}
S(E)_{a,b} = \delta_{a,b} - 2\pi i \delta(E-H_0) T_{a,b}(E) \,,
\end{eqnarray}
where ($a,b$) denote $MB$ channels, and 
the scattering T-matrix is
defined by the following coupled-channel equation
\begin{eqnarray}
T_{a,b}(E) = V_{a,b} + \sum_{c}V_{a,c} g_c(E) T_{c,b}(E) \,.
\end{eqnarray}
Here the meson-baryon propagator of channel $c$ is
\begin{eqnarray}
g_c(E) = < c \mid g(E) \mid c > \nonumber 
\end{eqnarray}
with 
\begin{eqnarray}
g(E)&=&\frac{1}{E-H_0 +i\epsilon} \, \nonumber \\
&=& g^P(E) -i\pi\delta(E-H_0)  \,,
\end{eqnarray}
where 
\begin{eqnarray}
g^P(E) &=& \frac{P}{E-H_0} \,.
\end{eqnarray}
Here $P$ denotes taking the principal-value part of any
integration over the propagator.
If $g(E)$  in Eq.(5) is replaced by $g^P(E)$, we then define
the K-matrix which is related to T-matrix by
\begin{eqnarray}
T_{a,b}(E)= K_{a,b}(E) - \sum_{c}T_{a,c}(E)[i\pi\delta(E-H_0)]_c K_{c,b}(E) \,.
\end{eqnarray}

By using the two potential formulation,
one can cast Eq.(5) into
\begin{eqnarray}
T_{a,b}(E)  &=&  t^{bg}_{a,b}(E) + t^{R}_{a,b}(E)  
\end{eqnarray}
with 
\begin{eqnarray}
t^{R}_{a,b}(E) &=& \sum_{N^*_i, N^*_j}
\bar{\Gamma}^\dagger_{N^*_i, a}(E) [G(E)]_{i,j}
\bar{\Gamma}_{N^*_j, b}(E)  \,.
\end{eqnarray}
The first term of Eq.(9) 
 is determined only by the non-resonant interaction 
\begin{eqnarray}
t^{bg}_{a,b}(E)= v^{bg}_{a,b} +\sum_{c} v^{bg}_{a,c}g_c(E) t^{bg}_{c,b}(E)\,.
\end{eqnarray}
The resonant amplitude Eq.(10) is determined by the
dressed vertex
\begin{eqnarray}
\bar{\Gamma}_{N^*,a}(E)  &=&
  { \Gamma_{N^*,a}} + \sum_{b} \Gamma_{N^*,b}
g_{b}(E)t^{bg}_{b,a}(E)\,,
\end{eqnarray}
and the dressed propagator
\begin{eqnarray}
[G(E)^{-1}]_{i,j}(E) = (E - M^0_{N^*_i})\delta_{i,j} - \Sigma_{i,j}(E) \,.
\end{eqnarray}
Here $M_{N^*}^0$ is the bare mass of the resonance state $N^*$, and
the self-energy is 
\begin{eqnarray}
\Sigma_{i,j}(E)= \sum_{a}\Gamma^\dagger_{N^*,a} g_{a}(E)
\bar{\Gamma}_{N^*_j,a}(E)\,.
\end{eqnarray}

Note that the meson-baryon propagator $g_a(E)$ 
for channels including an unstable particle, such as
$\pi \Delta$, $\rho N$ and $\sigma N$,
must be modified
to include a width.
 In the Hamiltonian formulation, this
amounts to the following replacement
\begin{eqnarray}
g_{a}(E) \rightarrow <a \mid\frac{1}{ E - H_0 - \Sigma_V(E)} \mid a> \,,
\end{eqnarray}
where the energy shift is
\begin{eqnarray}
\Sigma_V(E) = \sum_{i} \Gamma^+_V (i)\frac{P_{\pi\pi N}}{E-H_0 + i\epsilon}
\Gamma_V(i) \,.
\end{eqnarray}
Here $\Gamma_V$ describes the decay of $\rho$, $\sigma$ or $\Delta$
 in the quasi-particle channels. 
 
Eq.(5), Eqs.(9)-(16), and Eq.(8) are the starting points of our derivations.
From now on, we consider the formulation in the
partial-wave representation.
The channel labels, ($a,b,c$), will also include the usual
angular momentum and isospin quantum numbers.

\subsection{Unitary Isobar Model ($UIM$)}
\subsubsection{MAID}
The Unitary Isobar Model developed\cite{maid} by the Mainz group
is based on the on-shell relation Eq.(8).
By including only one hadron channel, $\pi N$ (or $\eta N$ ), Eq.(8) leads to
\begin{eqnarray}
T_{\pi N,\gamma N} = 
e^{i\delta_{\pi N}}cos\delta_{\pi N} K_{\pi N, \gamma N}\,.
\end{eqnarray}
where $\delta_{\pi N}$ is the pion-nucleon scattering phase shift.
By further assuming that  
$K=V=v^{bg} + v^R$, one can cast the above equation into the following form
\begin{eqnarray}
T_{\pi N,\gamma N}({\it UIM}) =
e^{\delta_{\pi N}}cos\delta_{\pi N} [v^{bg}_{\pi N, \gamma N}]
+\sum_{N^*_i}T^{N^*_i}_{\pi N, \gamma N}(E) \,.
\end{eqnarray}

The non-resonant term $v^{bg}$ in Eq.(18)  
is calculated from the standard Born terms but with an
 energy-dependent mixture of
pseudo-vector (PV) and pseudo-scalar (PS) $\pi NN$ coupling
and the  $\rho$ and $\omega$ exchanges. 
For resonant terms in Eq.(18), MAID
uses the following Walker's parameterization\cite{walk}
\begin{eqnarray}
T^{N^*_i}_{\pi N, \gamma N}(E) =f^i_{\pi N}(E)
\frac{\Gamma_{tot}M_i e^{i\Phi}}{M^2_i-E^2 - i M_i\Gamma^{tot}}
f^i_{\gamma N}(E)\bar{A}^i \,,
\end{eqnarray}
where $f^i_{\pi N}(E)$ and $f^i_{\gamma N}(E)$ are the form factors describing
the decays of $N^*$, $\Gamma_{tot}$ is the total decay width,
$\bar{A}^i$ is the $\gamma N \rightarrow N^*$ excitation strength.
The phase $\Phi$ is required by the unitary condition and is determined by
 an assumption  relating the phase of the total production amplitude to  the 
$\pi N$ phase shift and inelasticity.

\subsubsection{JLab/Yeveran UIM}

The Jlab/Yerevan UIM\cite{azn2} is similar to MAID.
But it implements the
Regge parameterization in calculating the 
amplitudes at high energies. It also uses a different procedure to
unitarize the amplitudes.

\vspace{1cm}

Both MAID and JLab/Yeveran UIM have been applied extensively to analyze the
data of $\pi$ and $\eta$ production reactions. Very useful new information
on $N^*$ have been extracted.

\subsection{Multi-channel K-matrix models}
\subsubsection{SAID}

The model employed in SAID\cite{said} is based on the on-shell
relation Eq.(8) with three channels:
$\gamma N$, $\pi N$, and $\pi\Delta$ which represents all other open channels.
The solution of the resulting $3\times 3$ matrix equation can be written
as
\begin{eqnarray}
T_{\gamma N,\pi N}(SAID) = A_I(1 + iT_{\pi N,\pi N}) 
+ A_RT_{\pi N,\pi N}\,,
\end{eqnarray}
where 
\begin{eqnarray}
A_I &=& K_{\gamma N, \pi N} 
- \frac{K_{\gamma N ,\pi\Delta}K_{\pi N,\pi N}}
{K_{\pi N,\pi\Delta}} \,, \\
A_R&=&\frac{K_{\gamma N, \pi\Delta}}{K_{\pi N, \pi\Delta}}\,.
\end{eqnarray}
In actual analysis, they simply parameterize $A_I$ and $A_R$ as
\begin{eqnarray}
A_I&=& [v_{\gamma N, \pi N}^{bg}] 
+ \sum_{n=0}^{M} \bar{p}_n z Q_{l_\alpha +n}(z) \,, \\
A_R&=& \frac{m_\pi}{k_0}(\frac{q_0}{k_0})^{l_\alpha}
\sum_{n=0}^{N} p_n (\frac{E_\pi}{m_\pi})^n \,,
\end{eqnarray}
where $k_0$ and $q_0$ are the on-shell momenta for pion and photon respectively,
$z=\sqrt{k^2_0+4m_\pi^2}/k_0$,  
$Q_L(z)$ is the legendre polynomial of second kind, 
$E_\pi = E_\gamma -m_\pi(1+m_\pi/(2m_N))$, and $p_n$ and
$\bar{p}_n$ are free parameters.
SAID  calculates 
$v^{bg}_{\gamma N,\pi N}$ of Eq.(23)
from the standard PS Born term and $\rho$ and $\omega$ exchanges.
The empirical $\pi N$ amplitude $T_{\pi N,\pi N}$ needed to
evaluate Eq.(20) is also available in SAID. 

Once the parameters $\bar{p}_n$ and $p_n$ in Eqs.(23)-(24)
are determined,
the $N^*$ parameters are then extracted by
fitting the resulting amplitude $T_{\gamma N,\pi N}$ at energies near the
resonance position to 
a Breit-Wigner parameterization(similar to Eq.(19)).
Very extensive data of pion photoproduction have been analyzed by SAID.
The extension of SAID to also analyze pion electroproduction data 
is being pursued.
 
\subsubsection{Giessen Model}
The 
coupled-channel model developed by the Giessen group~\cite{giessen}
can be obtained from Eq.(8) by taking the approximation
$K = V$. This leads to a matrix equation involving only the
on-shell matrix elements of $V$
\begin{eqnarray}
T_{a,b}({\it Giessen}) =
\sum_{c}[(1+i V(E))^{-1}]_{a,c} V_{c,b}(E) \,.
\end{eqnarray}
The interaction $V=v^{bg} +v^R$ is evaluated from tree-diagrams  
of various effective lagrangians.
The form factors, coupling constants, and resonance parameters are adjusted to
fit both the $\pi N$ and $\gamma N$ reaction data. They  include up to 
5 channels in some fits, and have identified
several new $N^*$ states. But further confirmations are needed to establish
their findings conclusively. 

\subsubsection{KSU Model}
The Kent State University (KSU) model\cite{kent}
can be derived by noting that the non-resonant amplitude $t^{bg}$,
defined by a $hermitian$ $v^{bg}$ in Eq.(11),  define
a S-matrix with the following properties
\begin{eqnarray}
S^{bg}_{a,b}(E) &=& \delta_{a,b} - 2 \pi i \delta(E-H_0) t^{bg}_{a,b}(E) \\
&=&\sum_{c}\omega^{(+)T}_{a,c}(E)\omega^{(+)}_{c,b}(E) \,,
\end{eqnarray}
where the non-resonant scattering operator is
\begin{eqnarray}
\omega^{(+)}_{a,c}(E) &=& \delta_{a,c} + g_a(E) t^{bg}_{a,c}(E) \,.
\end{eqnarray}
With some derivations, the S-matrix
Eq.(4) and the scattering T-matrix defined by Eqs.(9)-(14) can then
be cast into
following form
\begin{eqnarray}
S_{a,b}(E) =\sum_{c,d} \omega^{(+)T}_{a,c}(E)  R_{c,d}(E)
\omega^{(+)}_{c,b}(E) \,,
\end{eqnarray}
with
\begin{eqnarray}
R_{c,d}(E)=\delta_{c,d} + 2 i T^R_{c,d}(E)\,. \\
\end{eqnarray}
Here we have defined
\begin{eqnarray}
T^R_{c,d}(E) = \sum_{i,j} \Gamma^\dagger_{N^*_i,c}(E)
[G(E)_{i,j}{ \Gamma }_{N^*_j,d}(E) \,.
\end{eqnarray}
The above set of equations is identical to that used
in the  KSU model of Ref.\cite{kent}.
In practice, the KSU model fits the data by parameterizing
 $T^R$ as a Breit-Wigner
resonant form $ T^R \sim x \Gamma/2/(E- M - i\Gamma/2)$
and setting $\omega^{(+)} = B = B_1B_2\cdot\cdot\cdot B_n$, where
$B_i =exp (iX\Delta_i)$ is a unitary matrix.

The KSU model has been applied to $\pi N$ reactions, including pion
photoproduction. It is now being extended to investigate $\bar{K}N$ reactions.

\subsection{The CMB Model}

A unitary multi-channel isobar model with analyticity
 was developed\cite{cmb} in 1970's by the
Carnegie-Mellon Berkeley(CMB) collaboration
for analyzing the $\pi N$ data.  
The CMB model can be derived by assuming that the non-resonant 
potential $v^{bg}$ is also of the separable form of $v^R$ of Eq.(3)
\begin{eqnarray}
v^{bg}_{a,b} = \frac{\Gamma^\dagger _{L,a} \Gamma_{L,b}}{E-M_L}
+\frac{\Gamma^\dagger _{H,a} \Gamma_{H,b}}{E-M_H}
\end{eqnarray}
The resulting coupled-channel equations are identical to Eqs.(9)-(16), except
 that $t^{bg}_{a,b}=0$ and 
the sum over $N^*_i$ is now extended to include these two
distance poles $L$ and $H$.

By changing the integration variables and adding a substraction term,
 Eq.(14) can lead to CMB's dispersion
relations
\begin{eqnarray}
\Sigma_{i,j}(s) &=& \sum_{c} \gamma_{i,c} \Phi_c(s) \gamma_{j,c} \,,
\\
Re [\Phi_c(s)] &=& Re [\Phi_c(s_0)]
+\frac{s-s_{th,c}}{\pi} \int_{s_{th}}^{\infty}
\frac{Im[\Phi_c(s^\prime)]}{(s^\prime - s)(s^\prime-s_0)} d s^\prime \,.
\end{eqnarray}
Thus CMB model is analytic in structure which marks its difference
with all K-matrix models described above.

 The CMB model has been revived
in recent years by the Zagreb group\cite{zagreb} and a 
Pittsburgh-ANL collaboration\cite{pittanl} to
 extract the $N^*$ parameters 
from fitting the recent empirical $\pi N$ and  $\gamma N$ reaction 
amplitudes. The resulting $N^*$ parameters have very significant differences
with what are listed by PDG in some partial waves.
 In particular, several important issues concerning the extraction of
 the $N^*$ parameters in $S_{11}$ channel have been analyzed in  detail.


\subsection{Dynamical Models}

\begin{center}
{\bf A. In the $\Delta$ region}
\end{center}

Keeping only one resonance $N^*=\Delta$ and two channels $a,b= \pi N, \gamma N$,
Eqs.(9)-(14) are reduced to what were developed in
the Sato-Lee (SL) model\cite{sl}.  In solving exactly Eqs.(9)-(14),
the non-resonant interactions $v^{bg}_{\pi N,\pi N}$ and
$v^{bg}_{\pi N,\gamma N}$  are derived from the standard PV
Born terms 
and $\rho$ and $\omega$ exchanges by using an unitary transformation method. 

In the Dubna-Mainz-Taiwan (DMT) model\cite{dmt}, they depart from the formulation Eqs.(9)-(14)
by using the Walker's
parameterization defined by Eq.(19) to describe the resonant term
$t^{R}$ of Eq.(9). Accordingly,
their definition of the non-resonant amplitude also differs from Eq.(11): 
$t^{bg}_{c,b}$ in the right-hand side of Eq.(11) is replaced by the
 full amplitude $T_{c,b}$.
Furthermore, they follow MAID to calculate the non-resonant interaction
$v^{bg}_{\pi N,\gamma N}$ from an energy-dependent mixture of PS and PV
Born terms. 

Extensive data of pion photoproduction and electroproduction in the $\Delta$ region
can be described by both the SL and DMT models. However, the extracted 
$\gamma N \rightarrow \Delta$ form factors, in particular their bare form factors,
are significantly different. 
 
\begin{center}
{\bf B. In the second and third resonance regions }
\end{center}

Eqs.(9)-(16) are used in a 2-$N^*$ and 3-channel ($\pi N$, $\eta N$,
and $\pi \Delta$)  study\cite{yosh} of $\pi N$ scattering in $S_{11}$ partial wave,
aiming at investigating how the quark-quark interaction in
the constituent quark model can be determined directly by using the reaction data.
Eqs.(9)-(16) are also the basis of 
examining the $N^*$ effects\cite{otl} and one-loop coupled-channel effects\cite{ohlee}
on $\omega$ meson photoproduction and the
coupled-channel effects on $K$ photoproduction\cite{chitab}. 

The coupled-channel study of both $\pi N$ scattering and $\gamma N \rightarrow \pi N$
in  $S_{11}$ channel by Chen et al\cite{chen} includes
$\pi N$, $\eta N$, and $\gamma N$ channels.
Their $\pi N$ scattering calculation is performed by using Eq.(5), which
is of course equivalent to  Eqs.(9)-(14).
In their $\gamma N \rightarrow \pi N$ calculation, they neglect 
the $\gamma N \rightarrow \eta N \rightarrow \pi N $ coupled-channel effect, 
and follow the procedure of the DMT model to 
 evaluate the resonant term in terms of
the Walker's parameterization (Eq.(19)).
  They find that four
$N^*$  are needed to fit the
empirical amplitudes in $S_{11}$ channel up to $W = 2$ GeV. 

A coupled-channel calculation based on Eq.(5) has been carried out
by J\"ulich group\cite{julich} for $\pi N$ scattering.
They are able to describe the $\pi N$ phase shifts
up to $W=1.9$ GeV by including $\pi N$, $\eta N$,
$\pi \Delta$, $\rho N$ and $\sigma N$ channels and 5 $N^*$ resonances :
$P_{33}(1232)$, $S_{11}(1535)$, 
$S_{11}(1530)$, $S_{11}(1650)$ and $D_{13}(1520)$. They find that the Roper resonance
$P_{11}(1440)$ is completely due to the meson-exchange coupled-channel effects.

A coupled
channel calculation based on Eq.(5) for both $\pi N$ scattering 
and $\gamma N \rightarrow \pi N$ up to $W=1.5$ GeV has
been reported by Fuda and Alarbi\cite{fuda}. 
 They include $\pi N$, $\gamma N$, $\eta N$, and $\pi \Delta$
channels and 4 $N^*$ resonances :
$P_{33}(1232)$, $P_{11}(1440)$, $S_{11}(1535)$,
 and $D_{13}(1520)$. The parameters are adjusted to fit the empirical multipole amplitudes in
a few low partial waves.

Much simpler coupled-channel calculations have been  performed by using 
separable interactions. In the model of Gross and Surya\cite{gros}, such separable 
interactions are from simplifying the meson-exchange mechanisms 
in Figs~\ref{fig:mechanism_1}.(a)-c) 
as a contact term
like Fig.~\ref{fig:mechanism_1}(d). They include only $\pi N$ and 
$\gamma N$ channels  and 3 resonances: $P_{33}(1232)$, $P_{11}(1440)$ and $D_{13}(1520)$,
and restrict their investigation up to 
$W < 1.5$ GeV. To account for the inelasticities in $P_{11}$ and $D_{13}$, the $N^*\rightarrow
\pi\Delta$ coupling is introduced in these two partial waves. The inelasticities in other
partial waves are neglected.

A similar separable simplification is also used in 
the chiral coupled-channel models\cite{kais1,oset1} for strange 
particle production. There the separable interactions
 are directly determined from
the leading contact terms of SU(3) effective chiral lagrangian and hence
only act on s-wave partial waves.
They are able to fit the total cross section data for various strange
particle production reaction channels without introducing resonance
states. It remains to be seen whether these models can be further improved to account for
higher partial waves which are definitely needed to give an accurate description of
the data even at energies near production thresholds.


\begin{figure}[t]
\vspace{25pt}
\begin{center}
\mbox{\epsfig{file=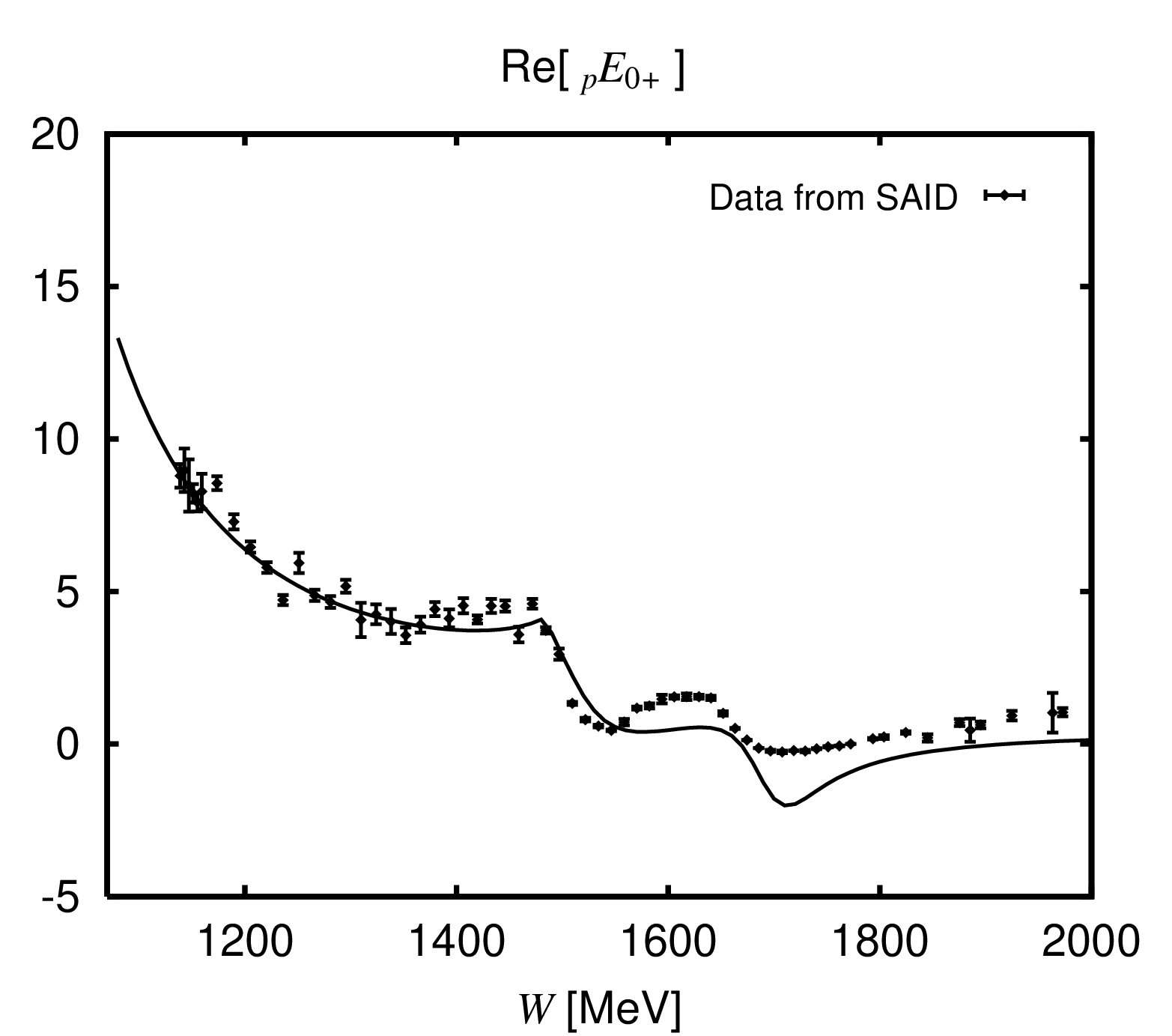,width=55mm}}
\mbox{\epsfig{file=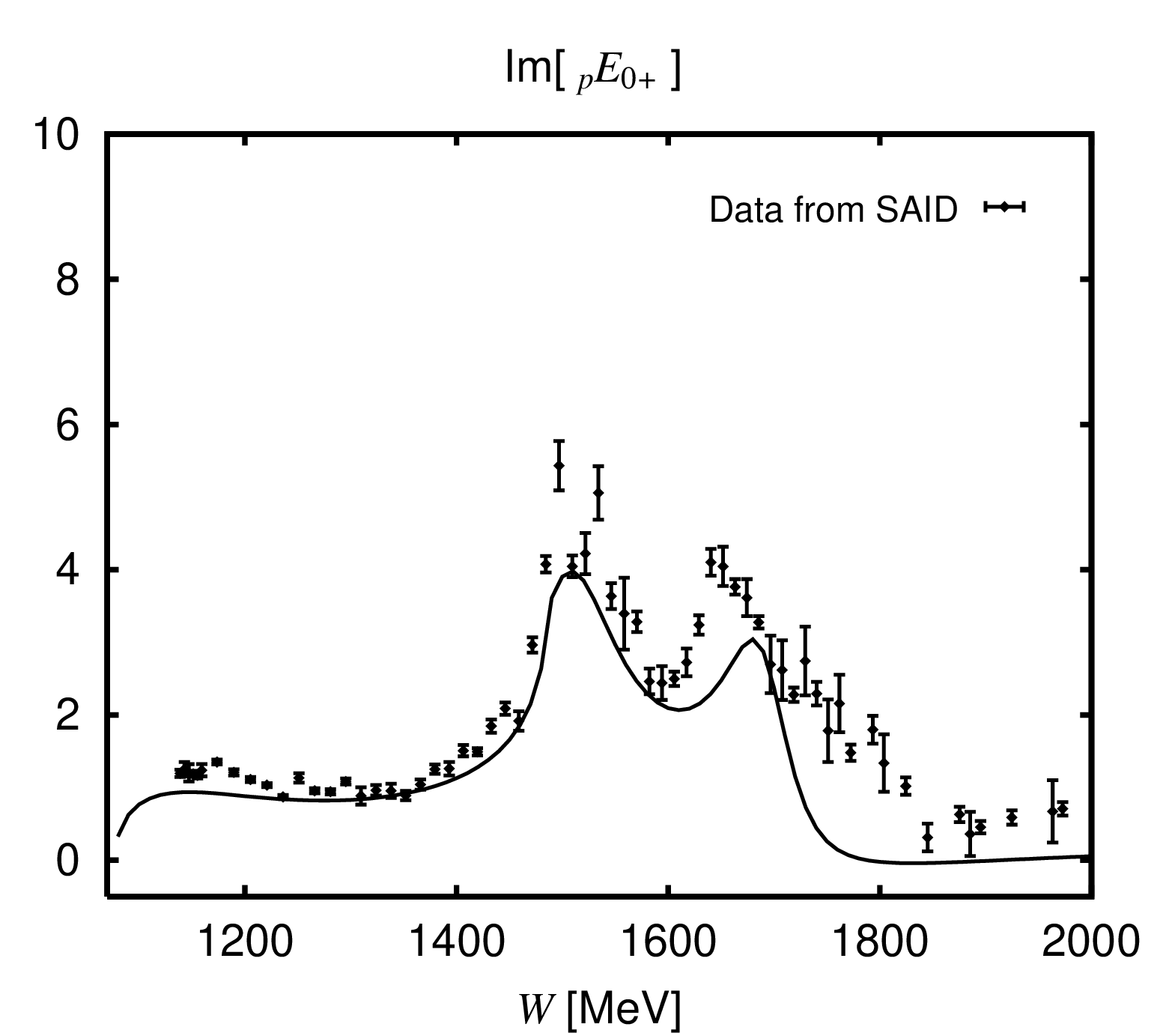, width=55mm}}
\end{center}
\caption{ }
\label{fig:e0}
\end{figure}

\section{Unitary $\pi\pi N$  Model}

All of the models described in section 2 rely on the assumption that the
$\pi\pi N$ $continuum$ can be expanded in terms of quasi-two-particle channels
such as $\pi\Delta$, $\sigma N$ , and $\rho N$.
These models are of course not satisfactory since they do not 
account for all of the effects due to the coupling with 
the $\pi\pi N$ channel.
It is necessary to develop a reaction model which also satisfies
the $\pi\pi N$ unitarity condition exactly, 
This can be done by extending the Hamiltonian Eqs.(1)-(3) to
include a vertex interaction $\Gamma_V$ to account for the
$\rho\rightarrow \pi\pi$ and  $\sigma\rightarrow \pi\pi$ decays  and to include
possible non-resonant $\pi\pi$ interaction 
$v_{\pi\pi}$. Such a formulation and numerical
methods for performing unitary calculations of
two-pion production cross sections are being pursued by Lee, Matsuyama,
and Sato (LMS)\cite{lms}. Here we only briefly describe this 
unitary $\pi\pi N$ model.

The coupled-channel equations from LMS  can also be cast into
 the form of Eqs.(9)-(16)
except that the driving term of Eq.(11) is replaced by
\begin{eqnarray}
v^{bg}_{a,b} \rightarrow \hat{V}_{a,b} = v^{bg}_{a,b}
+X_{a,b}(E)
\end{eqnarray}
with
\begin{eqnarray}
X_{a,b}(X) = Z_{a,b}(E) + \sum_{c} Z_{a,c}(E) g_c(E) X_{c,b}(E) \,.
\end{eqnarray}
The driving term of the above integral equation is
\begin{eqnarray}
Z_{a,b}(E)
&=&  \sum_{i\neq j} < a \mid \Gamma_V^+(i)
\frac{ P_{\pi \pi N}}
{E- H_0 - v_{\pi N,\pi N}- v_{\pi\pi} + i\epsilon}  \Gamma_V(j)
\mid b > 
\end{eqnarray}
Note that $i \neq j$ specifing the sum over $N^*$ states in the
 above equation is to avoid the double
counting of $\pi\pi N$ effect which is already included in the
dressed propagator defined by Eq.(15).

We have applied this unitary $\pi\pi N$ 
formulation to investigate $\pi N$ scattering 
and $\gamma N \rightarrow \pi N$
in $S_{11}$ channel up to $W=2$ GeV.
The channels included are $\pi N, \eta N, \pi \Delta$ and $\gamma N$.
The needed non-resonant interactions are generated from tree-diagrams 
Figs.1(a)-(d) using the unitary transformation method. Two $N^*$ states are 
included in the fits to the $\pi N$ scattering amplitude and
the $E_{0^+}$ amplitude
of $\gamma N \rightarrow \pi N$. Our results for $E_{0^+}$ amplitudes are
shown in Fig.2.
We see that we are not able to fit the data at $W > 1.63 $ GeV and
hence only the extracted $N^*(1535)$ parameters are reliable.
Our results are shown in the Table below and compared
 with the values from Chen et al\cite{chen} (DMT)
 and the quark model prediction of Capstick\cite{caps}.
It is interesting to note that LMS's bare value of the
$N^*(1535) \rightarrow \gamma N$ helicity amplitude $A_{1/2}$ is close to the
quark model prediction. Both the DMT and LMS predict that the meson cloud 
effect, the differences between the dressed values and bare values, 
is to reduce the bare values to the dressed values. This is rather
different from the situation  in the $\Delta$ region
where the meson cloud is to 
enhance the transition strength. The differences between DMT and LMS values
reflect their significant differences in calculating the coupled-channel effects.

\begin{table}
\begin{tabular}{llllllllll}

\hline
                &  & & & & &      \\
   &  $M_R$    & $\Gamma_R$ & &$\frac{\Gamma_\pi}{\Gamma_R}(\%)$ & & $A_{1/2}$ \\ \hline
DMT$^{16}$ & $1528\pm 1$ & $95 \pm 5 $ & & $40\pm 1$ &&  $81\pm 3$ (dressed)\\
                &  & & &    &&  $108\pm 4$ (bare)    \\
\hline

LMS$^{22}$  &1538&122&&36& &61.24 (dressed)\\
                &  & &  &   &  &  77.64 (bare)   \\
Quark Model$^{23}$ & & & & & &  76 \\
                \hline
\end{tabular}
\label{tab:nstar}
\end{table}

\begin{figure}[t]
\vspace{25pt}
\begin{center}
\mbox{\psfig{file=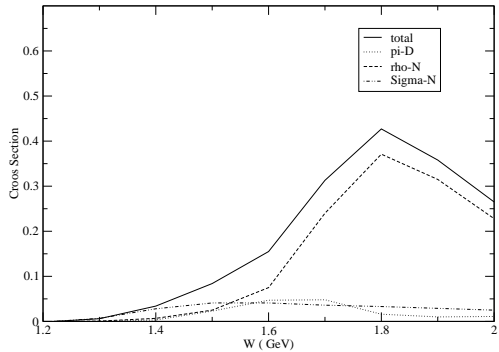, width=60mm, angle=-90}}
\end{center}
 \caption{ Calculated $\pi N \rightarrow \pi\pi N$ cross sections in
 $S_{11}$ channel. The partial cross sections through intermediate
$\pi \Delta$ (pi-D), $\rho N$ (rho-N) and $\sigma N$ (Sigma-N) are
also shown to compare with the coherent sum of these channels (Total), }
\label{fig:pin}
\end{figure}

To obtain reliable information for the second  $S_{11}$ resonance at about 1.6 GeV, we are
in the process of including  $\rho N$ and $\sigma N$ channels. The importance of
these two channels can be examined in a unitary calculation 
of $\pi N \rightarrow \pi\pi N$ cross sections.
This is achieved by using  the Spline-function expansion method which was developed in
our previous investigations of $\pi NN$ problem.
Our results of the partial cross sections of $\pi N \rightarrow \pi\pi N$ in
$S_{11}$ channel are shown in Fig.3.  Clearly, $\rho N$ channel
must be included   
for a dynamical interpretation of the second  $N^*$  and
to establish whether there exists third or even fourth $N^*$ in this 
channel. Our approach is clearly different from the investigation of
Chen et al.\cite{chen} who include only $\pi N$ and $\eta N$ channels and the fits
to the data are achieved by including up to
four $N^*$.

\section{Summary}
We have given a unified derivation of most of the models
for electromagnetic meson production reactions in the nucleon resonance region.
An extension of the coupled-channel approach to include $\pi\pi N$ channel
is briefly described and some preliminary results for
the $N^*(1535)$ excitation have been presented. 
Our complete calculations will be published elsewhere\cite{lms}.

\section*{Acknowledgments}
 This work is support in part by U.S. Department of Energy, Office
of Nuclear Physics, under Contract No. W-31-109-ENG-38, and
in part by Japan Soceity for Promotion of Science, Grand-in-Aid for Scientific
Research (C) 15540275.

%

\end{document}